\documentclass{mn2e}
\usepackage{psfig}
  
\def\solmas{{M$_\odot$}}
\def\simless{\mathbin{\lower 3pt\hbox
   {$\rlap{\raise 5pt\hbox{$\char'074$}}\mathchar"7218$}}}   
\def\simgreat{\mathbin{\lower 3pt\hbox
   {$\rlap{\raise 5pt\hbox{$\char'076$}}\mathchar"7218$}}}   
\def\etal{{\rm et al.}}

\def\solmas{{M$_\odot$}}
\def\solm{{M_\odot}}

\def\au{ AU}

\def\tff {t_{\rm ff}}

\def\AaA{{ A\&A}}
\def\AJ{{AJ}}
\def\ApJ{{ApJ}}
\def\apj{{ApJ}}
\def\apjl{{ApJL}}
\def\ApJL{{ApJL}}

\def\mnras{{MNRAS}}

\def\ARAA{{ARA\&A}}

  \makeatletter
  \ifx\CUP@mtlplain@loaded\undefined  
  \makeatother  
    
  \else
    
  \fi

\loadboldmathitalic 
\loadboldgreek      
  
\makeatletter
\ifx\CUP@mtlplain@loaded\undefined  
  \newfont\bit{cmbxti10 at 9pt}
\else
  \newfont\bit{mtbxti10 at 9pt}
\fi
\makeatother

\def\LaTeX{L\kern-.36em\raise.3ex\hbox{a}\kern-.15em
    T\kern-.1667em\lower.7ex\hbox{E}\kern-.125emX}

\newcommand{\gsim}{\mathrel{\hbox{\rlap{\lower.55ex \hbox {$\sim$}}
                   \kern-.3em \raise.4ex \hbox{$>$}}}}
\newcommand{\lsim}{\mathrel{\hbox{\rlap{\lower.55ex \hbox {$\sim$}}
                   \kern-.3em \raise.4ex \hbox{$<$}}}}

\title[Hierarchical cluster formation] {The hierarchical formation of a stellar cluster}
\author[I. A. Bonnell \etal]
  {Ian A. Bonnell$^1$\thanks{E-mail: iab1@st-and.ac.uk}, Matthew R. Bate$^2$ \& Stephen G. Vine$^1$\\ 
$^1$ School of Physics and
  Astronomy, University of St Andrews, North Haugh, St Andrews, Fife,
  KY16 9SS. \\
$^2$ School of Physics, University of Exeter, Stocker
  Road, Exeter EX4 4QL \\ }

\begin{document}

\maketitle

\begin{abstract}

Recent surveys of star forming regions have shown that most stars, and
probably all massive stars, are born in dense stellar clusters. The
mechanism by which a molecular cloud fragments to form several hundred
to thousands of individual stars has remained elusive. Here, we use a
numerical simulation to follow the fragmentation of a turbulent
molecular cloud and the subsequent formation and early evolution of a
stellar cluster containing more than 400 stars.  We show that the
stellar cluster forms through the hierarchical fragmentation of a
turbulent molecular cloud. This leads to the formation of many small
subclusters which interact and merge to form the final stellar
cluster.  The hierarchical nature of the cluster formation has serious
implications in terms of the properties of the new-born stars. The
higher number-density of stars in subclusters, compared to a more
uniform distribution arising from a monolithic formation, results in
closer and more frequent dynamical interactions. Such close
interactions can truncate circumstellar discs, harden existing
binaries, and potentially liberate a population of planets. We
estimate that at least one-third of all stars, and most massive stars,
suffer such disruptive interactions.
\end{abstract}

\begin{keywords}
stars: formation --  stars: luminosity function,
mass function -- globular clusters and associations: general.
\end{keywords}

\section{Introduction}
The advent of large, efficient infrared detectors has resulted in a
fundamental shift in our understanding of star formation. From the
older viewpoint that star formation was something done independently and in
isolation (Shu, Adams \& Lizano~1987), we now know that star formation
is a group activity whereby some tens to thousands of stars form
within a fraction of a parsec of each other (Clarke, Bonnell \&
Hillenbrand~2000).  In such environments, the forming stars interact
with each other on timescales comparable to their formation, resulting
in a highly dynamical picture of star formation (Bate, Bonnell \&
Bromm~2003; Bonnell \etal~1997; Chapman \etal~1992).  Infrared surveys of star forming
regions have repeatedly shown that most stars form in clusters (Lada
\etal~1991; Lada~1999; Clarke \etal~2000). This is found from both
unbiased and pointed observations of molecular clouds and it is
estimated that between 50 and 95 per cent of stars form in
clusters. Massive stars are even more likely to be found in young
stellar clusters (Clarke \etal~2000). Testi \etal (1997) used pointed
observations of Herbig AeBe stars and found a clear relation between
the star's mass and a surrounding stellar cluster. This implies 
a potential causal relationship between the cluster properties and
the mass of the most massive star.

The formation of stellar clusters has been an unsolved problem in
astronomy due to the intrinsic difficulty of fragmenting a molecular
cloud into a large number of stars, coupled with the numerical
difficulties in following the subsequent dynamical evolution.  The
spherical, nearly homogeneous nature of young stellar clusters has
generally been thought to imply that the preceding molecular cloud was
itself smooth and nearly spherical (Goodwin~1998). Fragmenting such an
object is extremely difficult as it requires the existence of
self-gravitating clumps that have gas densities significantly higher
than the mean gas density of the cloud (Bonnell~1999). 

In previous simulations involving smaller, lower-mass clouds,
fragmentation into many bodies occurs most readily when the cloud
contains filamentary structures which easily satisfy the above
condition (Bastien \etal~1991; Inutsuka \& Miyama~1997: Klessen,
Burkert \& Bate~1998; Bonnell~1999; Klessen~2000 \& Burkert~2000). Filamentary structures occupy a
small fraction of the total volume of the cloud. Thus, the free-fall
time of any self-gravitating perturbation in the structure is shorter
than the overall dynamical time of the system. Thus they can collapse
to form fragments before colliding together, forming binary and
multiple systems (Bonnell \etal~1991). A recent calculation (Bate
\etal~2003) modelled the formation of $\approx 50$ low-mass stars and
brown dwarfs from the fragmentation of a turbulent molecular cloud. In
this simulation (see also Klessen \etal~1998; Klessen \& Burkert~2000), the turbulence leads to
filamentary structures which then fragment as stated above. Cloud-cloud collisions have also been shown to lead to sheet-like and then filamentary structures
which subsequently fragment into multiple systems (Chapman \etal~1992).

In this paper, we present the results from the first numerical
simulation to follow the formation and evolution of a cluster
containing more than 400 stars. The primary new insight from this
simulation is that the cluster forms through a hierarchical process
of many small sub-clusters which grow and merge through the subsequent dynamics
to form the much larger final cluster. In Section 2 we detail
the calculation. In  Section 3 we describe the evolution of the forming
cluster while in section 4 we discuss the resultant mass distribution.
Section 5 describes the implications of a hierarchical cluster formation
process for stellar properties.
Our conclusions are given in section 6.

\section{Calculations}

We use a high resolution Smoothed Particle Hydrodynamics (SPH)
(Monaghan~1992) simulation to follow the fragmentation of an initially
uniform density molecular cloud containing $1000 \solm$ in a
diameter of $1$ pc  and a
gas temperature of $10$ K. SPH is a Lagrangian particle-based method
that calculates fluid properties by interpolation.  Calculations of
gravitational forces are facilitated using a tree-structure (Benz
\etal~1990). We use $5 \times 10^5$ individual SPH particles to follow
the gas dynamics.  We model the supersonic turbulent motions that are
observed to be present in molecular clouds by including a
divergence-free random Gaussian velocity field with a power spectrum
$P(k) \propto k^{-4}$ where $k$ is the wavenumber of the velocity
perturbations (Ostriker, Stone \& Gammie~2001). In three dimensions,
this matches the observed variation with size of the velocity
dispersion found in molecular clouds (Larson~1981). The velocities are
normalised to make the kinetic energy equal to the absolute
magnitude of the potential energy so that the cloud is marginally
unbound.  In contrast, the thermal energy is initially only 1 per cent
of the kinetic energy. Thus the cloud contains 1000 thermal Jeans masses,
\begin{equation}
\label{Jeans_mass}
M_J = \left(\frac{5 R_g T}{2 G \mu}\right)^{3/2} \left(\frac{4}{3} \pi
\rho \right)^{-1/2},
\end{equation}
where $T$ is the gas temperature, $R_g$ is the gas constant, $G$ is the gravitational
constant, $\mu$ is the mean molecular weight and $\rho$ is the density of the gas. Thus, in the absence of the turbulence and hence kinetic support, the
cloud could be expected to fragment into of order 1000 stars should
sufficient structure be present 
(Bonnell~1999). We force the
gas to remain isothermal throughout the simulation, emulating the
effect of efficient radiative cooling at the low gas densities
present. We do not include any feedback (radiative or kinematic) from
the newly formed stars. We expect that feedback, especially from
massive stars, will start to become important by the end of the
simulation.  The simulation was carried out on the United Kingdom's
Astrophysical Fluids Facility (UKAFF), a 128 CPU SGI Origin 3800
supercomputer.

Once fragmentation has produce a protostar, we replace the constituent
SPH particles with a single sink-particle (Bate, Bonnell \&
Price~1995).  These sink-particles, used to follow the newly formed
stars, interact only through gravitational forces and by accretion of
gas particles that fall into their sink-radii. Sink-particle creation
occurs when the densest gas particle (at a given time) and its $\approx 50$
neighbours are self-gravitating, subvirial and occupy a region smaller
than the sink-radius (200 \au). For the simulation reported here, this
requires a gas density of $\simgreat 1.5 \times 10^{-15}$ g cm$^{-3}$.
Gas particles that fall within a sink-radius of 200 \au\ are accreted
if they are bound to the sink-particle whereas all gas particles that
fall within 40 \au\ are accreted, regardless of their properties.  A
minimum number of SPH particles is required to resolve a fragmentation
event (Bate \& Burkert~1997) implying that our completeness limit for
fragmentation is approximately $0.1 \solm$.  Thus, 
we cannot resolve any protostars that would form with (initial) masses below
this limit.  Therefore, although we do resolve the bulk of the
fragmentation, there will be a number of lower-mass stars and brown
dwarfs that we do not capture in our simulation. Gas accretion onto
the stars then increases their masses and removes SPH particles from
the simulation. In order to minimise computational expense, we smooth
the gravitational forces between stars at distances of 160 \au. This
means that only the widest binaries and multiple systems are resolved
and that stellar collisions are not included in the simulation. The
use of gravitational softenning allows us to evolve the system further
than has previously been achieved and thus evaluate the formation
process of the stellar cluster.

\section{Hierarchical cluster formation}

The initial evolution of the molecular cloud is due to the turbulent
motions present in the gas. The supersonic turbulence leads to the
development of shocks in the gas, producing filamentary
structures (Bate, Bonnell \& Bromm~2003). The shocks also remove kinetic energy (assumed to be
radiated away) and thus remove the turbulent support
locally (Ostriker \etal~2001). The chaotic nature of the turbulence leads to
higher-density regions in the filamentary structures which, if they
become self-gravitating, collapse to form stars.

\begin{figure*}
\centerline{\psfig{figure=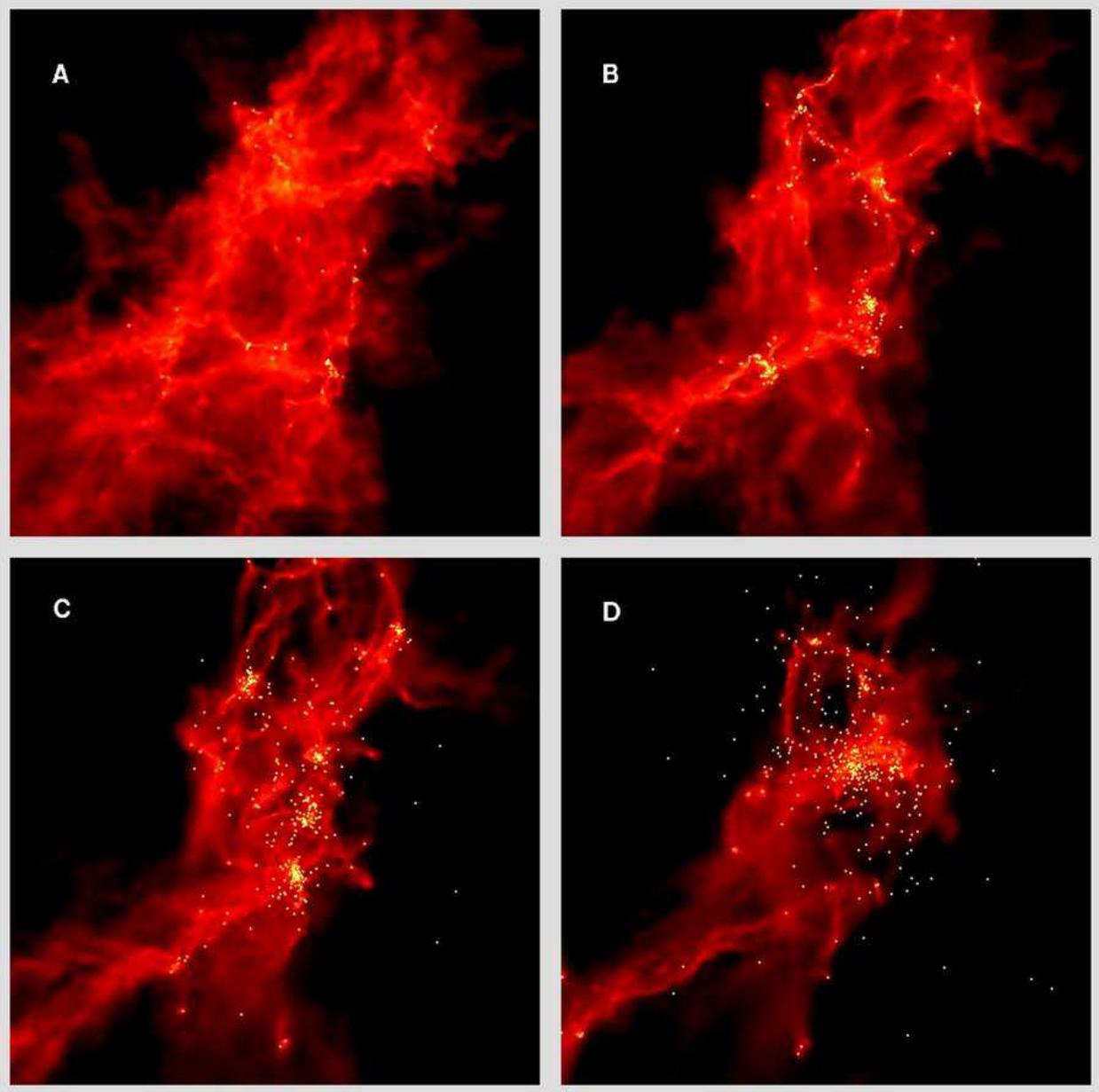,width=6.5truein,height=6.5truein}}
\caption{\label{cluform} The stellar cluster forms through the hierarchical
fragmentation of a turbulent molecular cloud. Each panel shows a
region 1 parsec on a side.  The logarithm of the column density is
plotted from a minimum of $0.025$ (black) to a maximum of 250 (white)
g cm$^{-3}$. The stars are indicated by the white dots. The four
panels capture the evolution of the 1000 \solmas\ system at times of
$1.0$, $1.4$, $1.8$ and $2.4$ initial free-fall times, where the
free-fall time for the cloud is $\tff = 1.9 \times 10^5$ years.  The
turbulence causes shocks to form in the molecular cloud, dissipating
kinetic energy and producing filamentary structure which fragment to
form dense cores and individual stars (panel A). The stars fall
towards local potential minima and hence form subclusters (panel B). These subclusters evolve by accreting more stars and gas, ejecting stars, and by
mergers with other subclusters (panel C).  The final state of the
simulation is a single, centrally condensed cluster with little
substructure (panel D). The cluster contains more than 400 stars and has a
gas fraction of approximately 16 per cent. }
\end{figure*}

Star formation occurs simultaneously at several different locations in
the cloud. Figure~\ref{cluform} plots the column density through the
molecular cloud at four different times over the $4.5 \times 10^5$
years ($2.6$ free-fall times, $t_{\rm ff}$) that we follow the evolution. The stars that
form first are in the highest density gas where the dynamical
timescale is the shortest. This generally occurs in the deepest parts
of local potential minima.  Surrounding clumps with slightly lower gas
densities form additional stars.  Both the stars and the residual gas
are attracted by their mutual gravitational forces and fall toward
each other.  Gas dynamics dampen the infall velocities allowing the
systems to rapidly merge to form a high density subcluster containing
from five to several tens of stars. The number of stars in each
subcluster increases as further star formation occurs nearby, and
these stars fall into the existing potential wells. This process
repeats itself until several hundred stars are formed and mostly
contained in five subclusters. The further evolution is marked by a
decreasing star formation rate as the subclusters, aided by the
dissipative effects of their embedded gas, sink towards each other and
finally merge to form one single cluster containing over 400
stars. The final cluster is approximately spherical in shape with a
centrally condensed core as is observed in young stellar
clusters (Hillenbrand~1997; Lada~1999).

\begin{figure}
\centerline{\psfig{figure=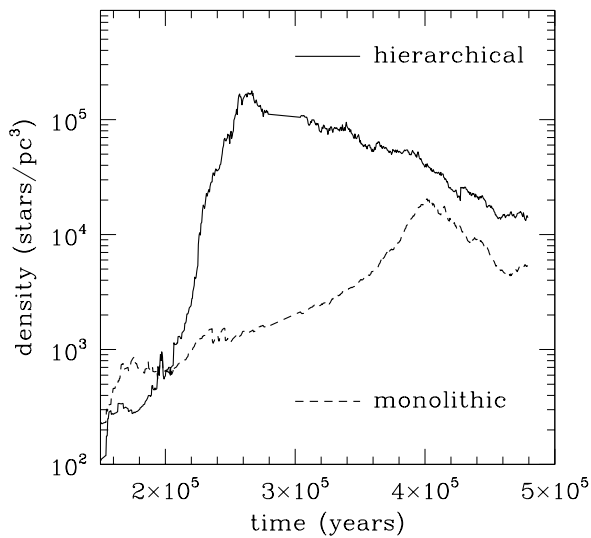,width=3.25truein,height=3.250truein}}
\caption{\label{numden} The evolution of the local and global stellar
number-densities are plotted as a function of time in years (the
initial free-fall time is $t_{\rm ff}=1.9 \times 10^5$ years). The
local stellar number-density (solid line) is calculated from the
volume required to hold the stars' 10 nearest neighbours; the star
having the median density is shown.  The global stellar number-density
(dashed line) is calculated from the volume required to contain fully
half of the total number of stars. This is equivalent to the local
density for a monolithic formation process (see text). The rapid rise
of the local (hierarchical) stellar density compared to the global
(monolithic) stellar density is due to the formation of subclusters.
The discrepancy between the two values decreases with time as the
substructure is erased to produce a single cluster. }
\end{figure}

The hierarchical nature of the formation process is illustrated in
figure~\ref{numden}, which shows the evolution of the local and global
stellar number-densities for the cluster. The local stellar density is
calculated for each star from the minimum volume required to contain
the star's ten nearest neighbours. We use the median value of this
distribution to quantify a typical local stellar density. In contrast,
the global stellar number-density is calculated from the volume
required to contain half of the total number of stars. This typifies
the stellar densities expected from a monolithic (homogeneous and
structureless) formation scenario.  We see that the local
number-density increases rapidly once the first stars form and fall
towards each other in their local subcluster.  The local number
density attains a maximum of $10^5$ stars pc$^{-3}$, up to 100
times that for a monolithic collapse.  The difference between the two
values indicates that the stars occupy but a small fraction of the
total volume of the star forming region.  This has significant
implications for the probability for interactions (see below).

 The local number-density decreases after reaching a maximum during
the subclustering phase. This decrease is due to the ejection of stars
from the subclusters through dynamical interactions, and due to the
kinetic heating during the merging of subclusters.  This
process erases the substructure fairly quickly, producing a single, centrally
condensed cluster.

\section{Accretion and the initial mass function}

\begin{figure}
\centerline{\psfig{figure=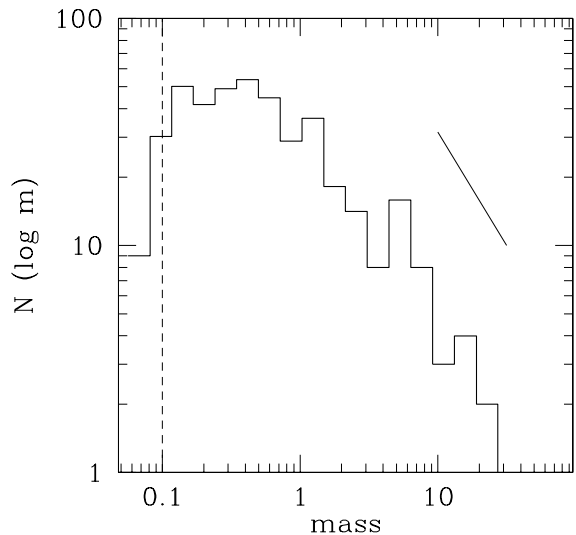,width=3.250truein,height=3.250truein}}
\caption{\label{clufimf} The final mass distribution of stars is plotted as a function of the logarithm of the mass. The distribution uses bins in the logarithm
of the mass such that a Salpeter IMF has a slope of $\Gamma=-1.35$. 
The minimum mass for the simulation is plotted as the dashed line while
the diagonal line notes a $\Gamma=-1$ slope. The higher-mass stars appear
to have a steeper distribution while intermediate-mass stars have a shallower one.
Stars below $\approx 1 \solm$ have an approximately flat distribution. }
\end{figure}.

In addition to acting as a reservoir for star formation and as a
damping force of the stellar dynamics, gas is accreted onto individual
stars, thereby increasing their masses. The stars compete for the gas,
with those in the bottom of their local potential wells accreting the
most and becoming the most massive stars (Bonnell
\etal~1997,~2001a). Thus, the first stars to form are frequently the
most massive due to this accretion process.  This process, termed
competitive accretion, is a leading candidate to explain the
apparently universal initial mass function (IMF) of stars (Bonnell
\etal~2001b; Klessen~2001). 

Here, gas accretion results in final stellar masses that range from
approximately $0.07$ to 27 \solmas. The final mass distribution, shown
in figure~\ref{clufimf} is consistent with observed IMFs
(Hillenbrand~1997; Luhman \etal~2000; Meyer \etal~2000). The
distribution has a near flat slope for low-mass stars, that turns into
an increasingly steeper slope for more massive stars (see also Bonnell
\etal~2001b; Bate \etal~2003). The higher-mass distribution is broadly
consistent with a $\Gamma\approx -1$ slope ($dN(log m) = m^{\Gamma}
dm$, where the Salpeter slope has $\Gamma=-1.35$) although could also
be fit by a shallower slope for intermediate-mass stars and a steeper
slope for high-mass stars. This high-mass slope is similar to the
recent result where high-mass stars are formed through a combination
of gas accretion and stellar mergers (Bonnell \& Bate~2002). In this
simulation, the gravitational potential wass not softened and binary
systems, formed through three-body capture, had separations as
small as 10 \au.

The median and mean stellar mass are $0.43$ and 1.38 \solmas,
respectively.  At the end of the simulation, 42 per cent of the total
mass remains in gas, although much of the gas is no longer bound to
the cluster due to the initial turbulence. The cluster contains 494
\solmas\ in a 0.25 pc radius, but only 16 per cent is in the form of
gas. This star formation efficiency, although comparable to that
observed for young stellar clusters (Lada~1991; Lada \& Lada~1995;
Clarke \etal~2000), does not include any background molecular cloud
not directly involved in the cluster formation process.  Furthermore,
as the fraction of mass in stars is an ever increasing function of
time, its value depends on when we halt the simulation.  At some point
in the process, feedback from the more massive stars is expected to
clear the remaining gas from the system. This gas expulsion will force
the cluster to expand as may be occurring in the ONC (Kroupa, Hurley
\& Aarseth~2001). Feedback from massive stars could also affect the
accretion process.  As feedback acts relatively quickly and only once
the massive stars have formed, its most probable effect will be a
freezing of the resultant mass function (Bonnell \etal~2001b).

\section{ Discussion}

The hierarchical nature of the formation process has many interesting
implications for star formation. Subclustering means that individual
stars are in regions of higher stellar number-density than they would
be for a monolithic formation process (Fig.~\ref{numden}; Fall \&
Rees~1985; Scally \& Clarke~2002).  The high number-density of stars,
coupled with the relatively small number of stars in each subcluster,
and thus smaller velocity dispersion, results in closer, and stronger
stellar interactions than would otherwise occur (Scally \&
Clarke~2002). Such stellar interactions can harden binaries to explain
the closest systems (Bate, Bonnell \& Bromm~2002), truncate
circumstellar discs (Hall, Clarke \& Pringle~1996; Bate \etal~2003)
decreasing their masses and thus lifetimes, trigger fragmentation in the
disc (Boffin \etal~1998; Watkins \etal~1998)  and possibly even liberate a population of
planets from their parent stars if planets form quickly (Bonnell \etal~2001c; Hurley \&
Shara~2002).  The maximum number-density of stars is sufficiently high
($10^7$ to $10^8$ stars pc$^{-3}$) that stellar mergers may
play a role in forming the most massive stars (Bonnell, Bate \&
Zinnecker~1998; Bonnell \& Bate~2002).

\begin{figure}
\centerline{\psfig{figure=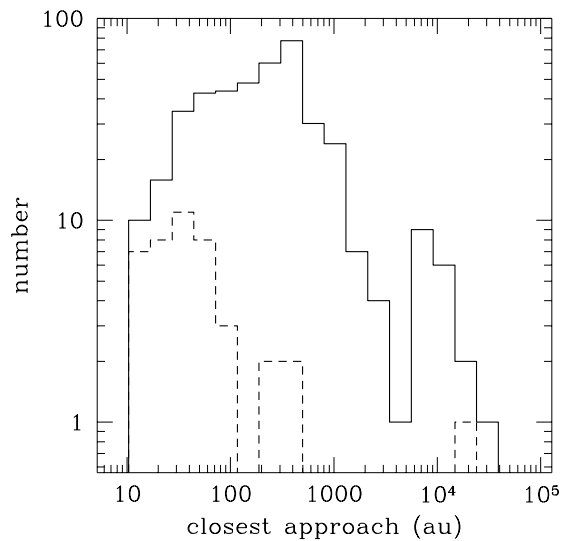,width=3.25truein,height=3.25truein}}
\caption{\label{closeapp} The distribution of minimum closest
approaches for all stars (solid line) and stars more than three
\solmas\ (dashed line) is plotted as a function of the logarithm of
the minimum separation. Gravitational softenning occurs for approaches
within 160 \au. Thus, the measured values for the closest approaches
are upper limits in these cases.}
\end{figure}

Figure~\ref{closeapp} plots the distribution of closest approaches
for each of the 418 stars formed in the simulation. 
This distribution is calculated, for each star, as the minimum distance to
any other passing star sometime  during the evolution. The distribution extends
from 10 \au\ to $> 10^4$ \au. The small peak at large separations indicates the
few stars that form in relative isolation in the molecular cloud, and never
enter into a cluster. 
Nearly half of
the stars in the simulation have interactions within the 160 \au\ 
resolution limit where we start to smooth the gravitational
forces. These are therefore upper limits for the actual
closest approaches. 

We see in figure~\ref{closeapp} that approximately one third (140/400) of the stars
have encountered another star within 100 \au, sufficiently close to
perturb the circumstellar disc (McCaughrean \& O'Dell~1996) or a binary system (Duquennoy \& Mayor~1991).  A
close passage of a star is likely to disrupt a circumstellar disc down
to one third of the minimum separation between stars (Hall \etal~1996), thus
limiting their mass reservoirs and lifetimes (although subsequent
accretion may replenish the discs [Bate \etal~2003]).  
Figure~\ref{closeapp} also
plots the corresponding distribution of closest approaches for more
massive stars, with $m \ge 3 \solm$. The great majority (34/40) of
these massive stars have had a close
interaction within 100 \au.  Interactions within 100 \au\ 
result in a disc truncated to 30 \au\ or less (Hall \etal~1996).
This suggests that many of the stars, and
virtually all the massive stars, found in young stellar clusters,
should have small (less than 30 \au) or non-existent discs.
A similar result was found in the smaller cluster 
simulation of Bate \etal~(2003) which resolved discs down to $\sim 10$ \au.

\section{Conclusions}

The results of the numerical simulation presented here point to a hierarchical
formation process for stellar clusters. Star formation occurs along filamentary
structures engendered in the molecular cloud due to supersonic turbulent motions. The stars fall together to form many small subclusters. These subclusters grow by accreting other stars and eventually merge to form a large-scale cluster
containing over 400 stars. The heirarchical nature of the formation process
means that close interactions between forming stars is much more important than would be in a monolithic formation process. Thus, stars are more likely
to have suffered close interactions that can truncate their circumstellar
discs, harden existing binaries and possibly liberate planets from their
parent stars. In particular, the hierarchical process and the corresponding 
higher stellar number-densities imply that approximately one third
of low-mass stars, and most high mass stars, should have had their discs truncated to within 30 \au.

Evidence for a hierarchical formation process exists in the
results of Testi \etal (2000) showing the hierarchical 
organisation of the sub-mm cores in the Serpens molecular cloud.
Detection of the  subclustered
phase at optical and near infrared wavelengths
is problematic due to the large dust extinction present in the cloud.
Fortunately, the upcoming Space Infrared Telescope Facility (SIRTF)
will observe star forming regions in the mid-infrared and should be
able to determine the level of hierarchical subclustering present
in even the youngest stellar clusters.

Finally, this numerical simulation demonstrates that there is a strong
link between the large-scale dynamics of star formation and the
small-scale stellar and protoplanetary disc properties. This
highlights the importance of a global approach to the problem. Future
studies will need to include even larger-scale processes such as the
formation and evolution of molecular clouds. In this way, we will be
able to determine how and when star formation is initiated and thus
tie star formation into Galactic evolution.

\section*{Acknowledgments}
We thank Mark McCaughrean, Keith Horne and the referee, Ant Whitworth
for many useful comments which
improved the paper. 
The computations reported here were performed using the U.K. Astrophysical
Fluids Facility (UKAFF).


\begin{thebibliography}{}


\bibitem[]{} Bastien P., Arcoragi J.-P., Benz W., Bonnell I., \& Martel H.,
1991, \ApJ, 378, 255

\bibitem[]{} Bate M. R., Bonnell I. A., Price N. M., 1995, \mnras, 277, 362

\bibitem[]{} Bate M. R., Burkert A., 1997, \mnras, 288, 1060

\bibitem[]{} Bate M. R., Bonnell I. A., Bromm V., 2002, \mnras, 336, 705

\bibitem[]{} Bate M.~R., Bonnell I.~A., Bromm V., 2003, \mnras, 339, 577

\bibitem[]{} Benz W., Bowers R. L., Cameron A. G. W., Press B., 1991, \apj, 348, 647

\bibitem[\protect\citename{Boffin} 1998]{1998MNRAS.300.1189B} Boffin
H.~M.~J., Watkins S.~J., Bhattal A.~S., Francis N., Whitworth A.~P., 1998,
MNRAS,  300, 1189

\bibitem[]{} Bonnell I. A., 1999, in {\sl The Origin of Stars and Planetary Systems}, (eds C.J. Lada and N. Kylafis) 479

\bibitem[]{} Bonnell I. A., Bate M. R., 2002, MNRAS, 336, 659

\bibitem[]{} Bonnell I. A., Bate M. R., Clarke C. J., Pringle J. E., 1997, \mnras, 285, 201

\bibitem[]{} Bonnell I. A., Bate M. R., Clarke C. J., Pringle J. E., 2001a, \mnras, 323, 785

\bibitem[]{} Bonnell I. A., Bate M. R., Zinnecker H., 1998, \mnras, 298, 93

\bibitem[]{} Bonnell I. A., Clarke C. J., Bate M. R., Pringle J. E., 2001b, \mnras, 324, 573

\bibitem[]{} Bonnell I. A., Martel H., Bastien P., Arcoragi J.-P., Benz W., 1991, \ApJ, 377, 553
 
\bibitem[]{} Bonnell I.~A., Smith K.~W., Davies M.~B., Horne K., 2001c, \mnras, 322, 859

\bibitem[\protect\citename{Chapman} 1992]{1992Natur.359..207C} Chapman S.,
Pongracic H., Disney M., Nelson A., Turner J., Whitworth A., 1992, Natur,
359, 207

\bibitem[]{} Clarke C.J., Bonnell I. A., Hillenbrand L. A., 2000, , in {Protostars and Planets IV} (eds V. Mannings, A. P. Boss and S. Russell), 151

\bibitem[]{} Duquennoy A., Mayor M., 1991, \AaA, 248, 485

\bibitem[]{} Fall S.~M., Rees M.~J., 1985, \ApJ, 298, 18

\bibitem[]{} Goodwin S. P., 1998, \mnras, 294, 47

\bibitem[]{} Hall S.M., Clarke C.J., Pringle J.E., 1996, \mnras, 278, 303

\bibitem[]{} Hillenbrand L. A., 1997, \AJ, 113, 1733

\bibitem[]{} Hurley J. R., Shara M. M., 2002, \ApJ, 565, 1251

\bibitem[]{} Inutsuka S., Miyama S.~M., 1997, \ApJ, 480, 681

\bibitem[\protect\citename{Klessen} 2001]{2001ApJ...556..837K} Klessen
R.~S., 2001, ApJ,  556, 837 

\bibitem[\protect\citename{Klessen} 2000]{2000ApJS..128..287K} Klessen
R.~S., Burkert A., 2000, ApJS,  128, 287

\bibitem[]{} Klessen R., Burkert A., Bate M. R., 1998, \ApJL, 501, L205

\bibitem[\protect\citename{Kroupa} 2001]{2001MNRAS.321..699K} Kroupa P.,
Aarseth S., Hurley J., 2001, MNRAS,  321, 699 

\bibitem[]{} Lada C., J. 1991, in {\sl The Physics of Star Formation and Early
        Stellar Evolution}, eds C. J. Lada, N. D. Kylafis, Kluwer, p. 329

\bibitem[]{} Lada E., 1999, {\sl The Origin of Stars and Planetary Systems} (eds C.J. Lada and N. Kylafis) 441

\bibitem[]{} Lada E., Evans N. J., Depoy D. L., Gatley I.,  1991, \ApJ, 371, 171

\bibitem[]{} Lada E.A., Lada C.J. 1995, \AJ, 109, 1682

\bibitem[]{} Larson R. B., 1981, \mnras, 194, 809

\bibitem[\protect\citename{Luhman} 2000]{2000ApJ...540.1016L} Luhman K.~L.,
Rieke G.~H., Young E.~T., Cotera A.~S., Chen H., Rieke M.~J., Schneider G.,
Thompson R.~I., 2000, ApJ,  540, 1016

\bibitem[]{} McCaughrean M., O'Dell R.,  \AJ, 1996, 111, 1977

\bibitem[]{} Meyer M. R., Adams F. C., Hillenbrand L. A., Carpenter J. M.,
Larson R. B., 2000, in {Protostars and Planets IV}
(eds V. Mannings, A. P. Boss and S. Russell). 121

\bibitem[]{} Monaghan J. J., 1992, \ARAA, 30, 543

\bibitem[]{} Ostriker E. C., Stone J. M., Gammie C. F.,  2001, \ApJ, 546, 980

\bibitem[]{} Scally A., Clarke, C., 2002, \mnras, 334, 156

\bibitem[]{} Shu F. H., Adams F. C., Lizano S., 1987, \ARAA, 25, 23

\bibitem[]{} Testi L., Palla F., Prusti T., Natta A., Maltagliati S., 1997, \AaA, 320, 159

\bibitem[]{} Testi L., Sarget A. I., Olmi L., Onello J. S., 2000, \apjl, 540, L53 

\bibitem[\protect\citename{Watkins} 1998]{1998MNRAS.300.1214W} Watkins
S.~J., Bhattal A.~S., Boffin H.~M.~J., Francis N., Whitworth A.~P., 1998,
MNRAS,  300, 1214



\end{thebibliography}
\end{document}